\documentclass[aps,prl,twocolumn,superscriptaddress]{revtex4-1}

\usepackage{graphicx}
\usepackage{bm}
\usepackage{amsmath}
\usepackage{amsfonts}

\newcommand{\beq}{\begin{equation}}
\newcommand{\eeq}{\end{equation}}
\newcommand{\beqa}{\begin{eqnarray}}
\newcommand{\eeqa}{\end{eqnarray}}

\begin{document}

% Title and abstract

\title{Bottom-up superconducting and Josephson junction devices \\ inside a Group-IV semiconductor}

 \author{Yun-Pil Shim}
 \email{ypshim@lps.umd.edu}
 \affiliation{Laboratory for Physical Sciences, College Park, Maryland 20740, USA}
 \affiliation{Department of Physics, University of Maryland, College Park, Maryland 20742, USA}
 \author{Charles Tahan}
 \email{charlie@tahan.com}
 \affiliation{Laboratory for Physical Sciences, College Park, Maryland 20740, USA}
 \date{\today}

% Abstract
\begin{abstract}
Superconducting circuits are exceptionally flexible, enabling many different devices from sensors to quantum computers. Separately, epitaxial semiconductor devices such as spin qubits in silicon offer more limited device variation but extraordinary quantum properties for a solid-state system. It might be possible to merge the two approaches, making single-crystal superconducting devices out of a semiconductor by utilizing the latest atomistic fabrication techniques. Here we propose superconducting devices made from precision hole-doped regions within a silicon (or germanium) single crystal. We analyze the properties of this superconducting semiconductor and show that practical superconducting wires, Josephson tunnel junctions or weak links, superconducting quantum interference devices (SQUIDs), and qubits are feasible. This work motivates the pursuit of ``bottom-up'' superconductivity for improved or fundamentally different technology and physics.
\end{abstract}

%\pacs{}
\maketitle

% Main body
%\section{Introduction}
%{\it Introduction}---

The Nb/AlO$_{\mathrm x}$/Nb (or Al/AlO$_{\mathrm x}$/Al) Josephson junction (JJ) has become almost ubiquitous for superconducting (SC) applications such as magnetometers \cite{Clarke_jltp1976}, voltage standards \cite{Niemeyer_apl1984,Benz_apl1996}, logic \cite{Likharev_ieee1991}, and qubits \cite{SC_qubit_review}. 
This follows a long history of development beginning in force with the IBM Josephson digital computer program during the 1970's \cite{anacker_IBM1980}, which pioneered the tunneling JJ technology (mostly based on Pb-alloy tunnel junctions, the critical current spread and the instability of the Pb limited its applicability). Nb-based tunnel junctions such as Nb/Al-AlO$_{\mathrm x}$/Nb \cite{gurvitch_apl1983} and Nb/AlO$_{\mathrm x}$/Nb \cite{ketchen_apl1991,nakagawa_ieee_as1991} proved to be more reliable and stable and have become the material of choice in many traditional SC devices while Al/AlO$_{\mathrm x}$/Al junctions are typically preferred for  quantum computing at milli-Kelvin operating temperatures. 

But heterogeneous devices such as these can pose problems, especially for low-power or quantum applications, where losses in or at the interfaces of the various materials (e.g, surface oxides on the superconductor, JJ insulator, substrate, interlayer dielectrics) can limit device quality dramatically. 
Possible solutions include better materials \cite{zkim_prb2008,pappas1,pappas2}, weak-link junctions \cite{vijay_prl2009}, symmetry protection \cite{gladchenko_natphysics2009}, or 3D cavity qubits \cite{paik_prl2011}. 
Here we consider an alternative approach: atomically-precise \cite{fuechsle_nnano2012}, hole-doped SC silicon \cite{bustarret_nature2006} (or germanium \cite{herrmannsdorfer_prl2009}) JJ devices and qubits made entirely out of the same crystal. Like the Si spin qubit, our super-semi \cite{cohen_rmp1964} JJ devices exist inside the ``vacuum'' \cite{tyryshkin_nmat2012,Steger_science2012} of ultra-pure silicon, far away from any dirty interfaces. We predict the possibility of SC wires, JJs, and qubits, calculate their critical parameters, and find that most known SC qubits should be realizable. This approach may enable better devices and exotic SC circuits as well as a new physical testbed for superconductivity.

Our proposal builds off of experimental progress in three different areas.
First, the list of SC materials has expanded to include doped covalent semiconductors \cite{blase_nmat2009},
particularly Si \cite{bustarret_nature2006,skrotzki_apl2010} and Ge \cite{herrmannsdorfer_prl2009,skrotzki_ltp2011}.
Extremely high doping rates (of acceptors) above the equilibrium solubility were achieved by gas-immersion laser doping (GILD) or ion implantation and annealing, and SC was observed in these high density hole systems.
Second, rapid progress in precise and high-density doping (of donors) in Si \cite{fuechsle_nnano2012} and Ge \cite{scappucci_nanolett2011} utilizing atomic layer doping and scanning tunneling microscope (STM) lithography has opened a new world of possible semiconductor devices, including single dopant qubits \cite{kane_nature1998}, single-atom-wide wires \cite{weber_science2012}, and even vertically-stacked 3D nanodevices \cite{scappucci_nanotech2011,klesse_apl2013,mckibbin_nanotech2013}. 
These same techniques should be applicable to acceptor incorporation. 
Finally, SC and Si/Ge qubits are widely considered to be leading candidates for fault-tolerant quantum computing (QC); yet both have negatives that combination may improve. For example, coherence times in isotopically enriched and chemically purified Si can reach seconds \cite{tyryshkin_nmat2012}, while SC qubits offer a huge range of design-space due to their macroscopic nature. Motivated by these results, we consider the following questions: What are the relevant properties of hole-doped SC ``wires'' in Si? What is required to create properly-placed, hole-doped SC Si Josephson junctions? And if such fabrication requirements are plausible, would such devices be of interest for qubits or other JJ circuits?
The answers to these questions are not obvious {\em a priori} given this unusual SC semiconductor system.

\section{Results}

\begin{table*}
\caption{\label{tab:sc_parameters} Superconductivity in hole-doped group IV materials and conventional metals. Only the parameters observed or estimated in the references cited were shown. All are type II superconductors except for Al and Pb. For type II superconductors, $H_{\mathrm c}$ is upper critical field for doped semiconductors, and lower critical field for Nb.}
\begin{ruledtabular}
\begin{tabular}{llllllll}
 material & $n_{\mathrm h}$(cm${}^{-3}$)  & $T_{\mathrm c}$(K) & $H_{\mathrm c}$(T) & $\xi_0$(nm) & $\xi$(nm) & $\lambda_{\mathrm L}$(nm) & $\lambda$(nm) \\
 \hline
 C:B \cite{ekimov_nature2004} & 4.6$\times$10$^{21}$ & 4 & 3.4 & - & 10 & - & -   \\
 C:B \cite{takano_drm2007} & 8.4$\times$10$^{21}$ & 11.4 & 10.8 & - & 5.51 & - & - \\
 Si:B \cite{marcenat_prb2010} & 4$\times$10$^{21}$ & 0.6 & 0.1 & 1000 & - & 60 & - \\
 Si:B \footnote{estimated values in the main text.} & 4$\times$10$^{21}$ & 0.6 & 0.1 & 1300 & 57 & 36 & 650 \\
 Si:Ga \cite{skrotzki_apl2010}\footnote{The SC region is at the interface between Si and $\text{SiO}_2$.} & 1$\times$10$^{22}$\footnote{peak Ga density} & 7 & 9.4 & - & 6 & - & 3700 \\
 Ge:Ga \cite{herrmannsdorfer_prl2009} & 4.3$\times$10$^{20}$ & 0.45 & 0.3 & - & 33 & - & $\sim 10^5$ \\
 \hline
 Al \cite{marder_book} & - & 1.18 & 0.01 & 1,300-1,600 & - & 16-50 & - \\
 Pb \cite{marder_book}& - & 7.20 & 0.08 & 51-96 & - & 39-63 & - \\
 Nb \cite{marder_book}& - & 9.3 & 0.2 & 38 & - & 39 & -
\end{tabular}
\end{ruledtabular}
\end{table*}

\subsection{Superconductivity in silicon}

By doping a semiconductor or an insulator above the metal-insulator-transition density, it has been expected that the host material turns into a superconductor \cite{cohen_rmp1964}.
Superconductivity has been observed in many such materials. (See Ref. [\onlinecite{blase_nmat2009}] and [\onlinecite{Iakoubovskii_physicaC2009}] for reviews.)
Particularly, superconductivity in hole-doped, group-IV materials have been found in 
diamond \cite{ekimov_nature2004}, silicon \cite{bustarret_nature2006}, and germanium \cite{herrmannsdorfer_prl2009}.
Various methods, such as high-pressure high-temperature(HPHT) treatments \cite{ekimov_nature2004} and growth using chemical vapor deposition (CVD) \cite{takano_apl2004,takano_drm2007} for C:B, GILD \cite{bustarret_nature2006,marcenat_prb2010,dahlem_prb2010} for Si:B, 
ion implantation and annealing  for Si:Ga \cite{skrotzki_apl2010,fiedler_prb2011} and Ge:Ga \cite{herrmannsdorfer_prl2009,skrotzki_ltp2011,fielder_prb2012}, 
were used to achieve very high hole densities required for superconductivity.
Table \ref{tab:sc_parameters} summarizes the superconducting parameters of the hole-doped group IV materials including those calculated here. They are compared with the conventional metal superconductors.

Superconductivity in silicon was first reported in Ref.~\cite{bustarret_nature2006}, by heavily doping a Si layer with boron (B) (above its equilibrium solubility in Si, $6\times 10^{20}\text{cm}^{-3}$).
This led to the very high hole density of $n_{\mathrm h} \simeq 5\times 10^{21} \text{cm}^{-3}$ and superconductivity was observed below $T_{\mathrm c} \simeq 0.35$K, although the SC Si layer (thickness $\simeq$ 35nm) was inhomogeneous with long tails in the superconducting and diamagnetic transitions.
Later experiments \cite{marcenat_prb2010} with much more homogeneous samples (thickness $\simeq$ 80-90nm) allowed systematic measurements of the dependence of the superconductivity on system parameters, such as the density and the external magnetic field. 
The highest $T_{\mathrm c}\simeq$0.6K was observed for the B density $c_{\mathrm B}\simeq$8 at.\%. (1 at.\% means 1\% of Si atoms are replaced with B atoms, which corresponds to 5$\times 10^{20}\text{cm}^{-3}$.) and the minimum B density $c_{\mathrm c}$ for superconductivity was $c_{\mathrm c}\simeq$2 at.\%. 
The critical field $H_{\mathrm c2}$ for $c_{\mathrm B}$=8 at.\% was measured to be 0.1T.
The experimental results agree well with the conventional Bardeen-Cooper-Schrieffer (BCS) theory \cite{BCS} for superconductors of type II.

We estimate the characteristic parameters of this superconductor for $c_{\mathrm B}$=8 at.\%. 
The observed critical temperature $T_{\mathrm c}$=0.6K corresponds to a zero temperature energy gap $\Delta(0)$=1.76$k_{\mathrm B} T_{\mathrm c}$=91$\mu$eV.
The characteristic lengths in an ideal (pure and local) SC and the more realistic effective values from Ginzburg-Landau (GL) theory have the following relations \cite{Tinkham_book}: 
\beqa
\frac{\xi(T)}{\xi_0} &=& \frac{\pi}{2\sqrt{3}}\frac{H_{\mathrm c}(0)}{H_{\mathrm c}(T)} \frac{\lambda_{\mathrm L}(0)}{\lambda(T)} \label{eq:xi_xi0}\\
\lambda(T) &=& \lambda_{\mathrm L}(T) \left( 1 + \frac{\xi_0/l}{J(0,T)} \right)^{1/2} \label{eq:lambdaT}
\eeqa
where $\xi_0$ ($\xi$) is the BCS (GL) coherence length and  $\lambda_{\mathrm L}$ ($\lambda$) is the London (effective) penetration depth, respectively.
$l$ is the mean free path and $J(R,T)$ is a function of length $R$ and the temperature $T$ defined by BCS \cite{BCS}. 
Using $H_{\mathrm c2}(0)$=$\Phi_0/ 2\pi \xi^2(0)$ with $H_{\mathrm c2}(0)$=0.1T where $\Phi_0$=$h/2e$ is the flux quantum, we obtain the GL coherence length $\xi(0)\simeq$57nm.
The London penetration depth can be calculated as $\lambda_{\mathrm L}(0)$=$\sqrt{m_{\mathrm h}/\mu_0 e^2 n_{\mathrm h}}\simeq$36nm
with the hole density $n_{\mathrm h}\simeq 4\times 10^{21}\text{cm}^3$ and the heavy hole effective mass $m_{\mathrm h}\simeq 0.5m_{\mathrm e}$ with $m_{\mathrm e}$ being the bare electron mass.
Since the system with $l\simeq$ 3nm ($\ll \xi,\lambda$) is in the dirty limit, using equations (\ref{eq:xi_xi0}) and (\ref{eq:lambdaT}),
we obtain $\xi_0$=$12\xi^2(0)/\pi^2 l \simeq$1300nm and $\lambda(0)\simeq\lambda_{\mathrm L}(0)\left(\xi_0/l\right)^{1/2}\simeq$650nm.
The GL parameter $\kappa=\lambda/\xi\simeq 11$ is consistent with type II superconductivity.  
These characteristic lengths are comparable to conventional metallic superconductors.

% Figure1
\begin{figure*}
  \includegraphics[width=\linewidth]{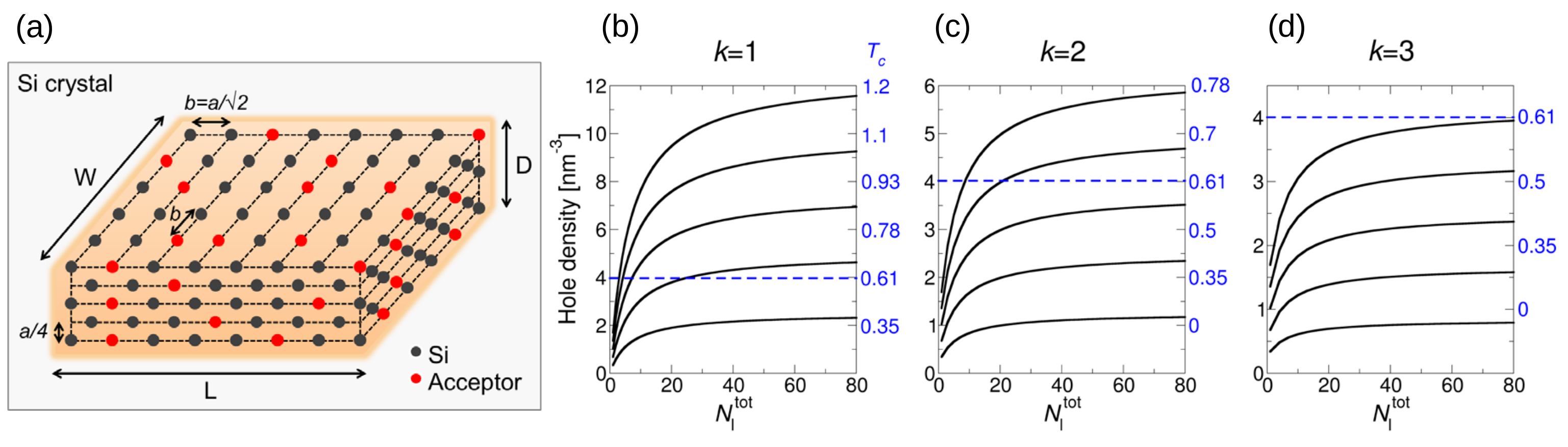}\\
  \caption{ {\bf Hole-doped SC silicon.} ({\bf a}), A specific region inside a single Si crystal, far from any noisy interfaces or surfaces, is hole-doped to sufficient acceptor density to go superconducting. The hole cloud, depicted by the orange region, has a larger extent than the lithographically doped region due to finite spread of the hole wave function. 
({\bf b-d}),  Hole density as a function of the number of total monolayers $N_{\mathrm l}^{\mathrm{tot}}$ of the doped region for different layer-doping-rates $r_{\mathrm D}$=  5, 10, 15, 20, 25 at.\% from bottom to top. Every layer ({\bf b}), every other layer ({\bf c}), and every 3rd layer ({\bf d}) is doped. The critical temperature $T_{\mathrm c}$
is shown on the right side of each plot. The blue dashed line indicates the density $n_{\mathrm h}$ = 4$\times 10^{21} \text{cm}^{-3}$ corresponding to the highest observed $T_{\mathrm c}$=0.6K for boron in silicon. 
    }
  \label{fig:hole_doping}
\end{figure*}

% STM doping
Superconducting devices in silicon such as JJs, superconducting quantum interference devices (SQUIDs), and SC qubits could be constructed out of hole-doped regions within the crystal. The doping method (GILD) used for demonstrating the highest $T_{\mathrm c}$ SC Si crystals so far \cite{marcenat_prb2010} may not be suitable for the epitaxially-encapsulated, nano-scale devices envisioned here.
Another method provides an alternative route: STM lithography has been used to precisely implant P dopants in Si.
STM lithography is a new technique that allows atomically precise doping of semiconductors.
We will briefly summarize the steps of P doping in Si.
A Si (001) surface with 2$\times$1 reconstruction (dimerization) is prepared and terminated with hydrogen resist. 
An STM tip is used to selectively remove some of the hydrogen atoms on the surface \cite{lyding_apl1994} (either across broad swaths of the crystal surface or down to single hydrogen atoms), exposing regions of unmasked silicon atoms.
A phosphine ($\text{PH}_3$) gas is introduced, which bonds selectively to the exposed silicon sites.
At least three adjacent desorbed dimers are needed for a P to replace a surface Si atom.
A phosphine molecule is chemisorbed to a dimer, dissociating into PH$_2$+H at room temperature.
Further annealing at 350$^{\circ}$ allows recombination/dissociation processes resulting in a P atom incorporated into the top Si layer ejecting a Si atom \cite{fuechsle_nnano2012}.  
In this way, P atoms were then incorporated into the exposed regions (via atomic layer doping), with positioning accuracy to one lattice site  \cite{fuechsle_nnano2010}.
The resulting 1D or 2D impurity sheet could reach very high doping rate, up to 1 in every 4 Si atoms being replaced with a P atom \cite{weber_science2012}.
It is not necessary to use an STM tip for the hydrogen desorption step, other lithographic techniques may be possible.
This process can be repeated to make stacked $\delta$-doped layers as was demonstrated in Ge \cite{scappucci_nanotech2011,klesse_apl2013} and Si \cite{mckibbin_nanotech2013}.

\subsection{Superconducting wires}
The SC Si:B realized by GILD was in a 2D layer with a thickness of tens of nanometers. For SC circuits and JJ applications, forming SC wires will be essential. We consider the use of atomic layer doping and STM lithography to dope B (or other acceptor) atoms into the Si crystal to achieve the very high hole density necessary for SC wires. 
Since this approach achieved a P density much higher than the B density reached in SC Si doped by GILD,  
higher hole doping rates may be possible (hence possibly higher critical temperatures) together with extremely fine control on the position and size of the SC region.    
Figure 1a shows a Si crystal doped with acceptor atoms. The lithographic region has length $L$, width $W$ and depth $D$.  
We assume that every $k$-th layer is doped with doping rate of $r_{\mathrm D}$.
If $N_{\mathrm l}$ monolayers are doped, the depth $D$=$(a/4)(N_{\mathrm l}-1)k$ and the total number of monolayers in the lithographic region is $N_{\mathrm l}^{\mathrm{tot}}$=$(N_{\mathrm l}-1)k+1$.
The total number of B dopants $N_{\mathrm D}$ is given by 
\begin{equation}
N_{\mathrm D} = \left( \frac{W}{b}+1 \right) \left( \frac{L}{b}+1 \right) r_{\mathrm D} N_{\mathrm l},
\end{equation}
where $b$=$a/\sqrt{2}$=3.84$\AA$, with $a$=$5.43\AA$ being the lattice constant of Si. 
To estimate hole density, we have to take into account the finite range of the holes \cite{weber_science2012}.
For the P impurities with Bohr radius 2.5nm, the effective electron density region has a diameter $d_{\mathrm B}$ ranging from 1 to 2 nm. 
An isolated B impurity in Si has a Bohr radius of 1.6nm \cite{gasseller_nanolett2011}, and we choose $d_{\mathrm B}$=1nm. 
Assuming all B dopants are activated, the hole density $n_{\mathrm h}$ is given by
\begin{equation}
n_{\mathrm h} = \frac{N_{\mathrm D}}{ \left( W + d_{\mathrm B} \right) \left( L + d_{\mathrm B} \right) \left( D + d_{\mathrm B}\right)}~.
\end{equation} 
For $W$ and $L$ much larger than $d_{\mathrm B}$, it is simplified as $n_{\mathrm h}$=$ (r_{\mathrm D} N_{\mathrm l}/b^2)/[(a/4)(N_{\mathrm l}-1)k+d_{\mathrm B}]$.
If the B density in a layer could reach the same level as the P in Si ($\sim$25 at.\%), 
the hole density of a single doped layer is 1.7$\times 10^{21}\text{cm}^{-3}$, which is above the critical hole density for superconductivity. In this case, using the experimentally observed density-dependence of the critical temperature, $T_{\mathrm c}$=$C \left(c_{\mathrm B}/c_{\mathrm c}-1\right)^{0.5}$ with $C\simeq 0.35$ \cite{marcenat_prb2010}, we obtain $T_{\mathrm c} \sim  0.3$K, but actual critical temperature could be lower than this due to the thin layer geometry \cite{grockowiak_sst2013}.
The maximum hole density is achieved for a thick doped region ($D \gg d_{\mathrm B}$) with every layer being doped ($k$=1). 
For $r_{\mathrm D}$=25 at.\%, $n_{\mathrm h}$=$1/ab^2$=$1.25\times 10^{22} \text{cm}^{-3}$ (a few times more than the highest density obtained by laser doping), we get a maximum $T_{\mathrm c} \simeq 1.2$K, which is comparable to the critical temperature of aluminium (Al). 
Although this could be possible if all the assumptions here are satisfied, to be more realistic all our calculations below will be for $T_{\mathrm c}$=0.6K which has been experimentally realized.
Figures\ 1b to 1d show the hole density as a function of depth $D$ when every layer ($k$=1), every other layer ($k$=2), or every third layer ($k$=3) is doped, respectively, for different doping rate $r_{\mathrm D}$. 
The highest observed $T_{\mathrm c}$ of 0.6K for Si:B should be reasonable for applications, {\it e.g.} quantum devices based on Al with $T_{\mathrm c} \sim$ 1K start to have problems due to quasiparticles at $T \sim$ 200mK,  and the requirements for classical applications such as photon detectors are much less restrictive.
To reach the $T_{\mathrm c}$=0.6K, we need at least three doped layers if each layer is maximally doped (25 at.\%.) for $k=1$, corresponding to the minimum depth of the lithographic layer $D$=$a/2$=0.27nm and the hole layer $D+d_{\mathrm B}$=1.27nm. The density strongly depends on the depth for small $D$ (i.e. small $N_{\mathrm l}$) and saturates to $4r_{\mathrm D}/(k a b^2)$ for large $D$. 
For thin SC wires, a cross section area larger than $10^3 \text{nm}^3$ is preferable (i.e. $W,D \gtrsim 30$nm) to prevent quantum phase slips \cite{tinkham_apl2002}.

% Figure2
\begin{figure}
  \includegraphics[width=\linewidth]{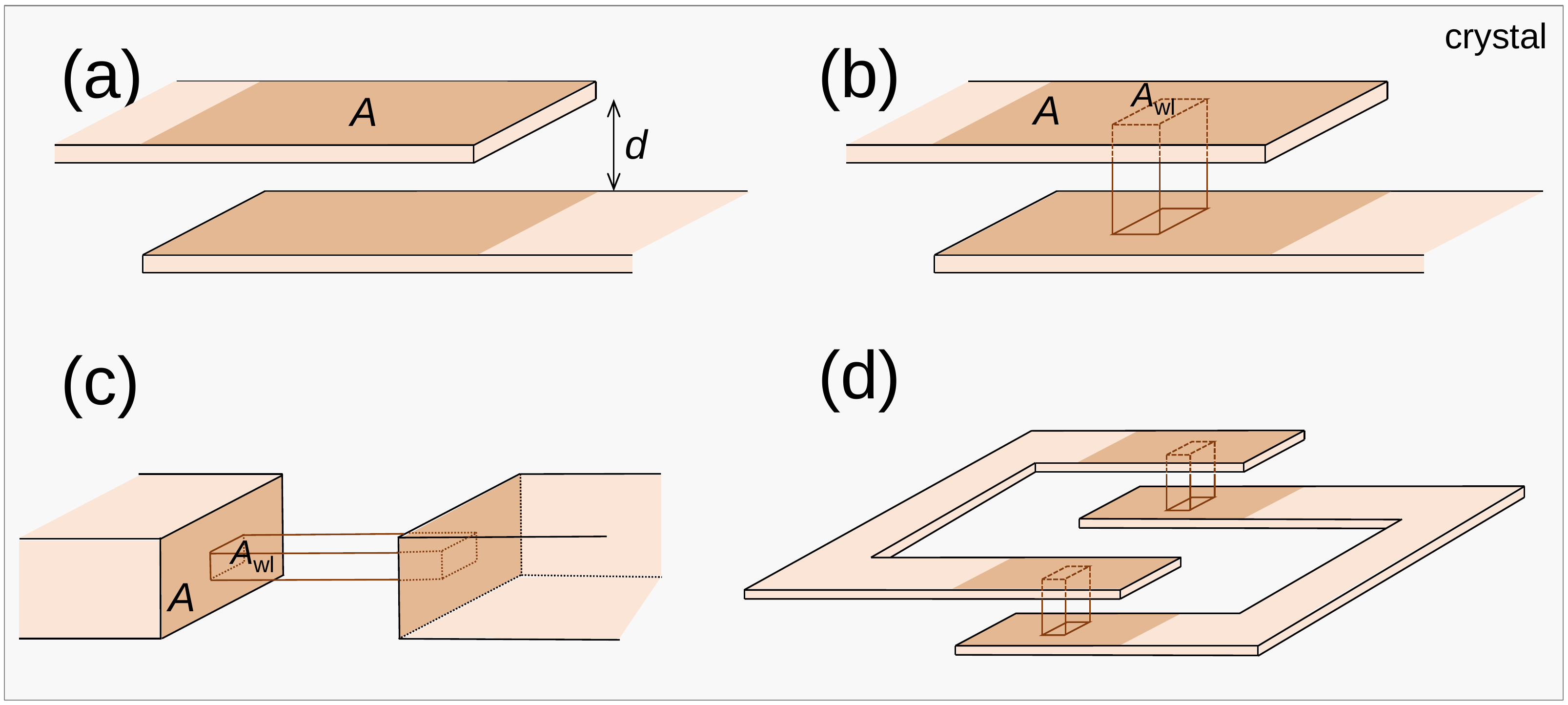}\\
  \caption{{\bf Super-semi JJs and SQUID geometries.} Examples of JJ devices that can be constructed inside the semiconductor are shown. Wire-figures depict the extent of the hole wave function. ({\bf a}), Superconducting tunnel junction (STJ) with overlapping area $A$ separated by distance $d$.
           ({\bf b}), Weak link JJ with overlapping SC layers. Critical current is determined by the bridge of cross section $A_{\mathrm{wl}}$ and length $d$, while the capacitance is determined by the overlap area $A$ and distance $d$. This geometry is suitable when a large overlap area $A$ (small charging energy $E_{\mathrm C}$) is required. 
           ({\bf c}), Weak link Josephson junction in a variable thickness bridge geometry (or STJ with no link),  suitable if large $A$ is not necessary. 
           ({\bf d}), SQUID circuit.}
  \label{fig:JJ}
\end{figure}

\subsection{Josephson junctions}
Josephson junctions are an essential ingredient for many SC applications.
Now we describe how one might realize a JJ made of this Si:B superconductor.
We will consider two types of JJ: the superconducting tunnel junction (STJ, Fig.\ 2a) and the weak-link \cite{likharev_weaklink_review} JJ (Figs.\ 2b and 2c). 
They are different in the way two superconductors are connected. A STJ consists of two superconducting electrodes divided by a tunneling barrier such as an insulating layer, while in weak-link JJs the two SCs are connected by a superconducting or metallic bridge. Traditionally, the STJ has been widely used due to its easier fabrication with a AlO$_{\mathrm{x}}$ barrier and its well-defined nonlinear current-phase relation. Weak-link junctions could be a good alternative especially in applications requiring high Josephson critical current and/or small size junction areas.

Two energy scales that characterize a JJ are the charging energy $E_{\mathrm C}$=$(2e)^2/2 C_{\mathrm{J}}$ for junction capacitance $C_{\mathrm{J}}$ and the junction energy $E_{\mathrm J}$=$\hbar I_{\mathrm{c}}/2e$ where $I_{\mathrm{c}}$ is the critical current (maximum DC Josephson current).
For the capacitance $C_{\mathrm{J}}$=$\varepsilon_{\mathrm{r}}\varepsilon_0 A/d$ with $\varepsilon_{\mathrm{r}}\simeq$ 12 for Si, the charging energy is given by
$E_{\mathrm C}$=3.0eV$\cdot$nm$\times d/A$.  
For the STJ, the critical current and the normal resistance $R_n$ has a relation \cite{ambegaokar_prl1963}
$I_{\mathrm{c}} R_{\mathrm{n}}$=$(\pi\Delta)/(2e) \tanh(\Delta/2k_{\mathrm B} T)$,
which reduces to $I_{\mathrm{c}} R_{\mathrm{n}}$=$\pi\Delta(0)/2e$ at zero temperature. Here $R_{\mathrm{n}}$ is the resistance of the junction in the normal state.
The above relation holds true for the weak link near $T$=$T_{\mathrm c}$, but at $T$=0,
$I_{\mathrm{c}} R_{\mathrm{n}}$=$1.32 \pi\Delta(0)/2e$ in the dirty limit \cite{kulik_omelyanchuk_1975}.
The junction energy at zero temperature then is $E_{\mathrm J}$=0.29 eV$\cdot\Omega\times(1/R_{\mathrm{n}})$ for the STJ and 0.39eV$\cdot\Omega\times(1/R_{\mathrm{n}})$ for the weak link.

% Figure3
\begin{figure}
  \includegraphics[width=0.8\linewidth]{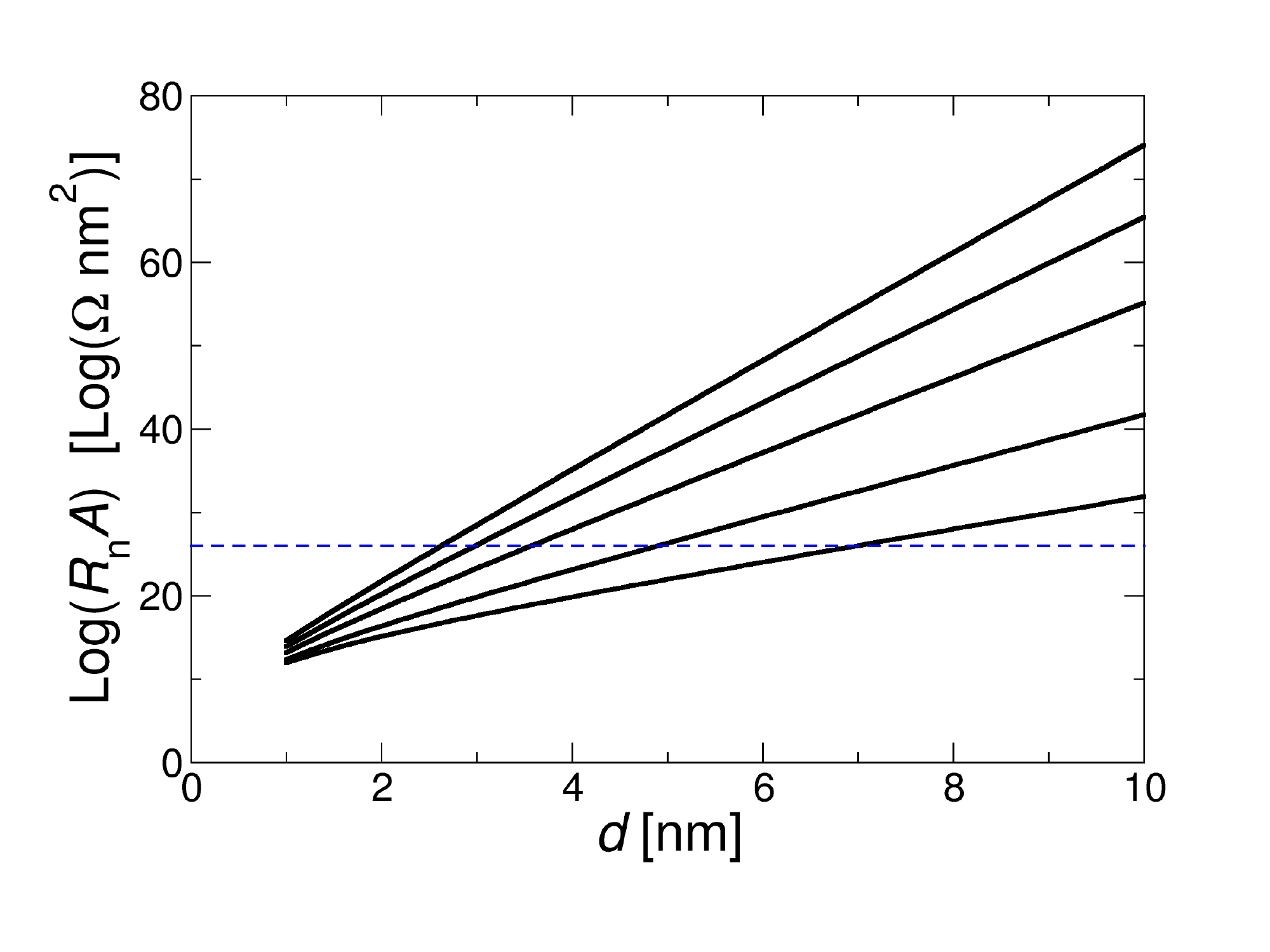}\\
  \caption{{\bf Tunneling resistance $R_{\mathrm{n}}$ as a function of the barrier width $d$.} Hole density is $n_{\mathrm h}$=4$\times 10^{21} \text{cm}^{-3}$, and barrier height $V_{\mathrm{b}}$ is, from top to bottom, $V_{\mathrm{b}}$=2.6, 2.4, 2.2, 2.0, 1.9 eV. The dotted blue line indicates a tunneling resistance corresponding to $V_{\mathrm{b}}$=2.4eV and $d$=3nm. 
  }
  \label{fig:tunnel_resistance}
\end{figure}
The normal resistance $R_{\mathrm{n}}$ has a simple form $\rho_{\mathrm{n}} d/A_{\mathrm{wl}}$ for a weak link where $\rho_{\mathrm{n}}$=$10^3\Omega\cdot\text{nm}$ \cite{marcenat_prb2010}. 
To estimate the normal resistance $R_{\mathrm{n}}$ of the tunnel JJ, we assumed a square potential barrier of width $d$ and height $V_{\mathrm{b}}$=$E_{\mathrm{g}}/2 + \varepsilon_{\mathrm{F}}$ where $E_{\mathrm{g}}$ is the energy gap of Si and $\varepsilon_{\mathrm{F}}$ is the Fermi energy of the holes for a given density. Then the tunneling conductance $G$ per unit area is given by
\beq
\frac{G}{A}=\frac{m_{\mathrm{h}}e^2}{2\pi^2\hbar^3} \int_0^{\varepsilon_{\mathrm{F}}} d\varepsilon_{\mathrm{z}} T(\varepsilon_{\mathrm{z}}) ~,
\eeq 
where $T(\varepsilon_{\mathrm{z}})$ is the transmission coefficient, 
\beq
T(\varepsilon_{\mathrm{z}}) = \frac{4 \varepsilon_{\mathrm{z}} \left( V_{\mathrm{b}} - \varepsilon_{\mathrm{z}} \right)}{ 4 \varepsilon_{\mathrm{z}} \left( V_{\mathrm{b}} - \varepsilon_{\mathrm{z}} \right) + V_{\mathrm{b}}^2 \sinh^2 \kappa d } ~,
\eeq
and $\kappa$=$\sqrt{\left( V_{\mathrm{b}} - \varepsilon_{\mathrm{z}} \right) 2 m_{\mathrm{h}}/\hbar^2}$.
Tunneling resistance $R_{\mathrm{n}}$=$1/G$.
We numerically calculated the tunneling resistance assuming the hole effective mass $m_{\mathrm{h}}$=0.5$m_{\mathrm{e}}$ and obtained $R_{\mathrm{n}} \simeq 10^4 e^{5.6d}/A [\Omega]$ with $d$ in unit of $\text{nm}$ and $A$ in unit of $\text{nm}^2$, for barrier height of $V_{\mathrm{b}}$=$E_{\mathrm{g}}/2 + \varepsilon_{\mathrm{F}} \simeq$ 2.4eV where $E_{\mathrm{g}}$ is the energy gap of Si and $\varepsilon_{\mathrm{F}}$ is the Fermi energy of holes.
Actually, the Fermi energy obtained by using the effective mass at low density is overestimated than the actual Fermi energy of high density holes \cite{bourgeois_apl2007}, but the barrier height $V_{\mathrm{b}}$ and shape could be significantly modified, e.g., by a spatially well separated heavily-doped region acting as a metallic gate. Therefore the resistance is tunable to a great extent. 
Figure \ref{fig:tunnel_resistance} gives the tunneling resistance as a function of the barrier width, for different barrier heights.
It clearly shows that $R_{\mathrm{n}}$ is proportional to $e^{\alpha d}/A$ for some constant $\alpha$. 
By tuning the barrier height, e.g. by lowering it, we can significantly relax the requirement on the necessary thinness of the barrier. 
If $d \lesssim$ 3nm was needed to obtain large enough tunneling current for $V_{\mathrm{b}}$=$E_{\mathrm{g}}/2 + \varepsilon_{\mathrm{F}} \simeq$ 2.4eV, we would need $d \lesssim$ 7nm for $V_{\mathrm{b}}$=1.9eV.

To overcome thermal fluctuations, the junction energy must be much larger than the temperature. In practice, $E_{\mathrm J} \gtrsim 5 k_{\mathrm B} T \approx 4.3 \mu\text{eV}$ for 10 mK.
The barrier distance $d$ of the STJ then must satisfy $d \lesssim 3$nm for $A$=1$\mu\text{m}^2$. 
The junction area $A$ cannot be much smaller since then the distance $d$ would need to be very small, 
but an external gate which could also be built of a separate doped region can control the tunneling barrier height and shape relaxing the restrictions at the cost of more complication in device design.  
A large junction area would be more easily implemented in the overlapping geometry of Fig.\ 2a, given that doping a thin layer with large area is probably easier than doping a small but thick region with STM lithography. 
For the weak link, on the other hand, the required condition is $A_{\mathrm{wl}}/d \gtrsim 0.01\text{nm}$ which could be easily satisfied, and the junction energy is independent of the total junction area $A$. Hence both Figs.\ 2b and 2c would be possible. 

If we want to avoid hysteresis in the $I-V$ curve as is usually required for dc SQUID application, we need an overdamped JJ and the junction quality factor $Q$=$\omega_{\mathrm{p}} R C_{\mathrm{J}}$ must be smaller than 1, where $\omega_{\mathrm{p}}$=$\sqrt{2 E_{\mathrm C} E_{\mathrm J}}/\hbar$ is the plasma frequency of the JJ.
$R$ is of order of $R_{\mathrm{n}}$ for the weak link and $R \sim R_{\mathrm{n}} e^{\Delta/k_{\mathrm B} T}$ for the STJ.
For the STJ to satisfy $Q<1$, typically a shunting resistance would be necessary to reduce the total resistance since $R$ is very large for an isolated tunnel junction (one would want to avoid this for quantum applications). 
Alternatively, SC-Insulator-Normal metal-Insulator-SC (SINIS) type junctions \cite{maezawa_apl1997} may be advantageous for achieving an overdamped JJ.
For the weak link, $Q$=5.5$\times 10^{-3}\sqrt{ A/A_{\mathrm{wl}}}$ and for $A_{\mathrm{wl}}$=100$\text{nm}^2$, $A<3.3\mu\text{m}^2$, allowing much smaller size than the STJ.
 
For a SQUID application such as shown in Fig.\ 2d, additional conditions should be satisfied to avoid magnetic hysteresis: 
$\beta_{\mathrm{m}}$=$2LI_{\mathrm{c}}/\Phi_0 < 1$ where $L$ is the inductance of the SQUID loop. 
STJs can easily satisfy this since the critical current is small, 
but a fairly large loop would be needed due to the large junction area $A \simeq 1\mu\text{m}^2$ required to overcome the thermal fluctuations as discussed above. 
On the other hand, weak link JJs open up the possibility of a nanoscale SQUID. 
For a square loop of area $1\mu\text{m} \times 1\mu\text{m}$, the geometrical inductance $L$ is  $\sim$ 3 pH for wire diameter of a few tens of nm, assuming the relative permeability of doped Si is 1 like most nonmagnetic metals. Then $\beta_{\mathrm{m}}  < 1$ translates into $A_{\mathrm{wl}}/d < 2\times 10^3$nm.
Typical values $A_{\mathrm{wl}}\simeq$ 100$\text{nm}^2$ and $d\simeq 10$nm would be suitable for a nano-SQUID.
Compared to the nano-SQUID based on the metallic SC bridges \cite{vijay_apl2010}, we could get much shorter weak links due to the much higher precision of STM lithography over e-beam lithography, allowing one to reach the short link limit with highly nonlinear inductance and larger modulation depth in critical current.

\subsection{Qubits}

% Figure4
\begin{figure}
  \includegraphics[width=\linewidth]{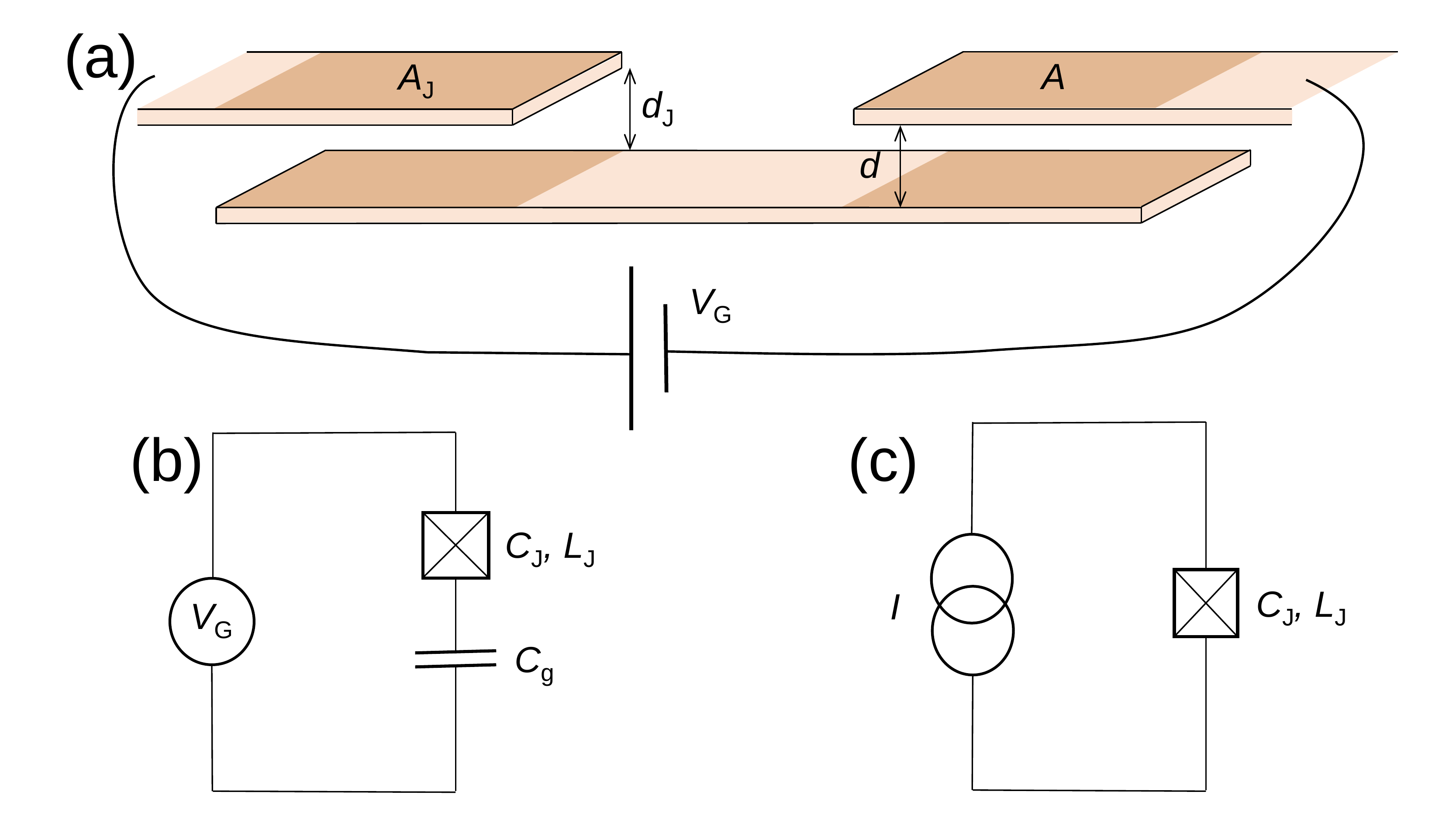}\\
  \caption{{\bf Superconducting qubit circuits.} ({\bf a}), Example charge qubit made of tunnel junctions. One tunneling junction $(A_{\mathrm{J}}, d_{\mathrm{J}})$ acts as the JJ and the other $(A,d)$ acts as just a capacitor by choosing different parameters. ({\bf b}), Equivalent circuit diagram of {\bf a}. ({\bf c}), Circuit diagram for phase qubit (current biased JJ), suitable for weak-link JJ.}
  \label{fig:qubit}
\end{figure}

Finally we consider the possibility of SC qubits in Si:B. 
The requirements on JJ parameters for qubits are different from the conditions for e.g. SQUID discussed in the previous section.
We will consider the core SC qubits---charge, phase, and flux---to estimate relevant parameters, 
noting that more complicated designs would relax the restrictions on the parameters significantly. 
A charge qubit is a single Cooper pair box connected to a Josephson junction where the two discrete low energy levels form a logical qubit space. 
Usually, a gate voltage is applied to tune the system to be in a sweet spot to reduce the effects of the charge noise, but in this case
its known sensitivity to charge noise might make a good probe of the charge environment of this system.
Figure~\ref{fig:qubit}a shows a possible geometry for a charge qubit and Fig.~\ref{fig:qubit}b is the equivalent circuit diagram.
By choosing different geometries for the two tunnel junctions e.g. $d_{\mathrm{J}} \lesssim$ 3nm and $d \gtrsim$ 10nm, the left junction can have large enough JJ energy to act as a JJ, while 
the right junction has negligible JJ energy and can be considered as a simple capacitor with capacitance $C_{\mathrm{g}}$.
The charge qubit is operated in a regime $k_{\mathrm B} T \ll E_{\mathrm J} \sim E_{\mathrm C} \ll \Delta$ where $E_{\mathrm C}$ is now the total charging energy $E_{\mathrm C}$=$(2e)^2/2(C_{\mathrm{J}}+C_{\mathrm{g}})$.
Assuming $T$=10 mK, a JJ with $d_{\mathrm{J}}$=2.5nm means that $A_{\mathrm{J}}$ should be $\simeq$ 1$\mu\text{m}^2$. 
The charging energy $E_{\mathrm C}$=3.0 eV$\cdot$nm $\times 1/(A_{\mathrm{J}}/d_{\mathrm{J}}+A/d)$ with $d_{\mathrm{J}}$=2.5nm and $A_{\mathrm{J}}$=1$\mu\text{m}^2$ constrains the geometry of the capacitor $A/d \ll 3.1\times 10^6$nm.
So we can choose, e.g., $A \sim 10^6\text{nm}^2$ and $d \sim$20nm.

High JJ critical current makes the phase qubit a good choice for the weak-link JJ.
Figure~\ref{fig:qubit}c shows a circuit diagram for a simple phase qubit.
A phase qubit operates in a regime with $k_{\mathrm B} T \ll E_{\mathrm C} \ll E_{\mathrm J}$, which translates into $d/A \gg 3\times 10^{-7}\text{nm}^{-1}$ and $A A_{\mathrm{wl}}/d^2 \gg 7.7\times 10^3\text{nm}^2$.
A reasonable set of parameters would be, e.g., $d \sim$ 10nm, $A_{\mathrm{wl}} \sim$ 100$\text{nm}^2$, and $A\sim 10^6\text{nm}^2$.
For the flux qubit, the simplest model is a loop with a JJ (rf-SQUID loop) coupled to an externally supplied flux. 
The flux qubit operates usually with $L_{\mathrm{J}} \lesssim L$, where $L_{\mathrm{J}}$=$\Phi_0/2\pi I_{\mathrm{c}}$. The loop inductance is relatively quite small compared to $L_{\mathrm{J}}$ for the typical geometries we have considered so far. This restriction can be lifted by using, e.g., a three JJ loop \cite{mooij_science1999,chiorescu_science2003}. 
More advanced qubits such as the transmon qubit \cite{transmon} are realizable by incorporating a big capacitor in the system, which is straight-forward.
In that case, the JJ can have a small junction area $A$ and both geometries in Figs.~\ref{fig:JJ}b and \ref{fig:JJ}c could be used.

\section{Discussion}
Our proposal is promising for new types of JJ devices. The noise environment of buried dopant layers has been reported to be quite low \cite{shamin_prb2011}, which is motivating for quantum applications, but obviously not sufficient. Fabrication requirements, as envisioned, have already been realized in the Si:P or Ge:P systems. Many JJ device and qubit geometries are possible beyond what are considered here, which may further reduce fabrication needs; lattice-site precision of impurities is not a fundamental requirement. An assumption in this work is the plausibility of acceptor placement with atomic-layer doping and STM lithography. B is currently being pursued in this context, but it is unproven whether the chemistry of adsorption and incorporation (e.g., of B$_2$H$_6$) will work in a similar manner as PH$_3$, nor whether the same densities can be achieved (1 in 4 atoms/ML). We have accounted for this by considering lower densities per monolayer. Quick B diffusion and clumping may limit further thermal anneal budgets, but this problem has already been overcome with low-temperature MBE \cite{weir_apl1994}. Local strain due to the strong B bonds is almost certainly present, but does not effect the epitaxial nature of the crystal \cite{bustarret_nature2006}. As potentially better dopant alternatives, Al (AlH$_3$) or Ga (GaH$_3$) for both Si and Ge should be pursued, as well as more advanced chemistry and surface preparation  approaches for STM lithography and doping (e.g., BCl$_3$ is used in GILD, so Cl might be considered instead of H).

The extension of the hole cloud ($\sim$ 1nm) would limit the sharpness of the SC region and hole density would drop to zero over this length. Since it is much smaller than the SC coherence length, the entire hole cloud is expected to be superconducting due to the proximity effect. One of the advantages of this single crystal device could be that there would be no Schottky barrier between heavily-doped (metallic) region and lightly-doped (semiconducting) region, and no interface states are expected in the interface between doped and undoped regions.

Stacked multilayer designs of electron doped Si devices were already demonstrated experimentally \cite{mckibbin_nanotech2013}. A second doped layer was grown on top of a nanowire capped by undoped Si of 50 $\sim$ 120 nm thick. The whole device was grown epitaxillay. The rather large separation between two doped layers in the experiment was needed to obtain smoother surface for the STM lithography of the top layer and also to get enough separation so that the top layer works as a metallic gate. On the contrary, we need the hole wavefunctions to overlap between layers for the 3D SC region. Thus we need much smaller separation between layers, and this could be a challenge. 
In fact, Ge may offer significant benefits over Si for JJ devices. Ge's clean surfaces and lower thermal requirements for good epitaxial growth \cite{scappucci_nanotech2011,klesse_apl2013} may allow for more and better 3D doped-layers as compared to Si (where the limits of epitaxial growth are more likely to result in surface roughness), with less diffusion due to thermal activation anneals. 

It is unclear what $T_{\mathrm c}$'s are possible in pure Ge (or Si) with other acceptors ($T_{\mathrm c}$'s of up to 7K \cite{skrotzki_apl2010} have been reported in Si:Ga/SiO$_2$ interface structures and even higher for diamond, and numerical simulation \cite{bourgeois_apl2007} suggests Al can lead to a higher critical temperature than B in Si). We have focused on Si due to the greater amount of experimental data versus density to guide our device proposals.  
Theory does not preclude electron-doped SC semiconductors \cite{cohen_rmp1964,bourgeois_apl2007}, but experimental efforts have so far shown no evidence \cite{grockowiak_sst2013}. 

The AlO$_{\mathrm{x}}$-based tunneling JJ has been very successful in many applications over the years, and other materials and different structures have also been studied for various devices \cite{golubov_rmp2004}. Building SC devices inside a semiconductor proposed here gives several advantages over conventional approaches. The availability of ultra-pure ${}^{28}$Si with less than 50 ppm ${}^{29}$Si \cite{becker_pssa2010} and the atomically precise positioning of dopants by STM lithography can help suppress the subgap states due to impurities in the JJ which is one of the main decoherence channels of SC qubits.
Flux noise is another possible source of decoherence for the SQUID \cite{anton_prl2013}, phase qubit \cite{bialczak_prl2007,sank_prl2012}, and flux qubit \cite{yoshihara_prl2006,kakuyanagi_prl2007}. It was suggested that the flux noise comes from the fluctuating spins at interfaces and surfaces of the device\cite{sendelbach_prl2008,sendelbach_prl2009}. 
Hyperfine interaction was proposed as a possible mechanism for the relaxation of the surface spins \cite{wu_prl2012}.
The lack of nuclear spin in enriched Si at the surface, the separation of the active device region from the surface, and the single crystal structure of the whole device should help significantly reduce the flux noise.
The epitaxially grown barrier in this proposal should also reduce any potential two-level fluctuators, as was shown for a crystalline Al$_2$O$_3$ barrier \cite{oh_sst2005,oh_prb2006,kline_sst2012}. 
In the Al/Al$_2$O$_3$/Al junctions, the critical current density showed a wider variation than amorphous AlO$_{\mathrm{x}}$ barriers. A qubit can be designed to be tolerable to these variations in junction critical current, but it involves a more complicated structure \cite{kline_sst2009}. 
For precisely positioned Si:B/Si/Si:B junctions, we expect less variations in system parameters for identically designed devices; barrier quality would be less important for weak-link JJs. Further, the devices can be constructed well below the surface of the semiconductor, away from oxide interfaces which typically cause loss. The quality of the superconducting semiconductor  itself is a new concern, and a good early experiment would be to determine the loss of such a device (e.g. via a cavity Q). Intriguing in these systems is the possibility of dissipation/quasiparticle engineering by manipulating the disorder of the implanted impurities.

Since the SC properties depend on the hole density, material parameters can be tuned with additional gates, allowing for a tunable SC-Normal Metal-SC (SNS) JJ \cite{doh_science2005}. In that case, the proximity effect will play an important role and needs to be fully taken into account, a topic also of interest in Majorana physics. In addition to these potentially improved material properties, STM lithography is suitable for small devices such as the nano-SQUID and allows for arbitrary 3D device designs for different types of qubits, detectors, circuits, etc.  

Progress in ``bottom-up'' fabrication techniques, such as STM lithography, have increased the space of devices worth pursuing. Our work further motivates the investigation of acceptor doping via precision techniques, beyond the context of single acceptor qubits \cite{rusko_prb2013} or for nanoscale but classical electronic devices. Successful demonstration of such proposed physics could enable not only the devices suggested in this work, but offer an atomically-configurable testbed for the nature and limits of semiconductor superconductivity (via, e.g., isotope variation, density, disorder, phonon, strain, and so on), for $T_{\mathrm c}$ engineering, as well as for new devices such as 3D SC device geometries, top-gated tunable JJs, or topological qubits \cite{nayak_rmp2008}.

%% Bibliography

%\bibliographystyle{nature}
%\bibliography{references}
%

%% Acknowledgement
\section{Acknowledgements} 
We thank R. Butera, M. Friesen, A. Mizel, B. Palmer, R. Ruskov for critical reading of the manuscript, and R. Joynt for useful conversations.

%% Author contributions
%\section{Author contributions}
%C.T. planned the project. Y.-P.S performed the theoretical and numerical calculations. 
%All the authors contributed to the interpretation of the results, discussions, and writing of the manuscript.
%
%
%% Additional information
%\section{Additional information}
%{\bf Competing financial interests:} The authors declare no competing financial interests.

\end{document}